\documentclass[%
reprint,
 amsmath,amssymb,
 aps,
]{revtex4-1}

\usepackage{graphicx}
\usepackage{dcolumn}
\usepackage{bm}

\begin{document}

\preprint{APS/123-QED}

\title{
 Topological and edge state properties of a three-band model for Sr$_2$RuO$_4$}%

\author{Yoshiki Imai}%
\email{imai@phy.saitama-u.ac.jp}%
\affiliation{Department of Physics, Saitama University, Saitama, 338-8570, Japan}

\author{Katsunori Wakabayashi}%
\affiliation{International Center for Materials Nanoarchitectonics (WPI-MANA), National Institute for Materials Science (NIMS), Tsukuba 305-0044, Japan}

\author{Manfred Sigrist}
\affiliation{Theoretische Physik, ETH-H\"onggerberg, CH-8093 Z\"urich, Switzerland}

\date{\today}

\begin{abstract}
Modeling the spin-triplet superconductor Sr$_2$RuO$_4$ through a three-orbital tight-binding model we investigate topological properties and edge states assuming chiral $p$-wave pairing. 
In concordance with experiments the three Fermi surfaces consist of two electron-like and one hole-like surface corresponding to the $\alpha$-, $\beta$- and $\gamma$-band on the level of a two-dimensional system. The quasi-particle spectra and other physical quantities of the superconducting phase are calculated by means of a self-consistent Bogoliubov-de Gennes approach for a ribbon shaped system. While a full quasiparticle excitation gap is realized in the bulk system, at the edges gapless states appear some of which have linear and others nearly flat dispersion around zero energy. This study shows the interplay between spin-orbit coupling induced spin currents, chiral edge currents and correlation driven surface magnetism. The topological nature of the chiral $p$-wave state manifests itself in the $\gamma$-band characterized by an integer Chern number. 
As the $ \gamma $-band is close to a Lifshitz transition in Sr$_2$RuO$_4$, changing the sign of the Chern number, the topological nature may be rather fragile.  
\begin{description}
\item[74.70.Pq,74.25.-q,73.20.-r]
\end{description}
\end{abstract}

\pacs{Valid PACS appear here}
\maketitle

\section{Introduction}
The layered perovskite compound Sr$_2$RuO$_4$ has attracted much interest for its unconventional superconductivity  appearing at $T_{\rm c}$$\sim$$1.5$ K in an essentially two-dimensional strongly correlated Fermi liquid~\cite{maen94,mack03}. 
NMR-Knight-shift measurements can be well interpreted assuming spin-triplet Cooper pairing~\cite{ishi98}. 
Enhanced zero-field relaxation in $\mu$SR experiments indicates the existence of the intrinsic spontaneous magnetic fields in the superconducting phase, suggesting broken time-reversal symmetry~\cite{luke98}. 
The leading candidate for the superconducting phase compatible with these two and several further experiments is the so-called chiral $p$-wave state, whose order parameter can be represented as 
\begin{eqnarray}
\bm d (\bm k)= \Delta_0 \hat{z}(k_x\pm {\rm i}k_y). 
\end{eqnarray}
 This state is the two-dimensional analog of A-phase (ABM state) of $^3$He superfluid~\cite{rice95}. It has a full energy gap and orbital angular momentum $L_z$ of the Cooper pairs along the $z$-axis (perpendicular to the basal plane). The two angular momentum states, $k_x+{\rm i}k_y$ and $k_x-{\rm i}k_y$, 
are degenerate and allow even for the formation of domains~\cite{volo85,mats99}.

The chirality of the pairing function leads to edge states, Andreev bound states, giving rise to spontaneous surface currents whose direction depends on the sign of $L_z$~\cite{volo85,mats99}. 
While the quasiparticle tunneling spectroscopy confirmed the subgap states near the surface~\cite{laube00,ying03,kashi11}, scanning Hall probes and the scanning SQUID microscopy experiments, however, did not confirm the existence of the supercurrent~\cite{tame03, kirt07,hicks10}. These studies have
been followed by discussion on the overall experimental consistency of the chiral $p$-wave phase for Sr$_2$RuO$_4$ (see e.g. Ref.~\cite{kallin12}). 

The chiral $p$-wave superconductivity has also attracted much interest for the topological nature of this phase. 
The chirality of the chiral $p$-wave superconducting state for the single band model is characterized by a topological number, which is defined as
\begin{eqnarray}
N=\frac{1}{4\pi}\int {\rm d}^2k\,\, \hat{\bm m} \cdot 
\left(
\frac{\partial \hat{\bm m}}{\partial k_x}\times 
\frac{\partial \hat{\bm m}}{\partial k_y}
\right), 
\end{eqnarray}
where $\bm m=({\rm Re}\Delta_{\bm k},-{\rm Im}\Delta_{\bm k},\varepsilon_{\bm k})$, and $\hat{\bm m}={\bm m}/{ |\bm m|}$~\cite{volo97}. $\Delta_{\bm k}$ ($\varepsilon_{\bm k}$) stands for the gap function (energy dispersion). This topological number is $+1$ or $-1$, which depends on the angular momentum of Cooper pairing~\cite{volo97,furu01}. The finite topological number is associated with the gapless chiral edge mode at the boundary of the domain wall with opposite chirality or at the surfaces.  

Sr$_2$RuO$_4$ has a K$_2$NiF$_4$-type lattice structure, isostructural with the parent compound of the high-$T_{\rm c}$ curate (La,Sr)$_2$CuO$_4$ and shows strong two-dimensional anisotropy.
The Fermi surfaces of cylindrical topology are formed by three bands, the $\alpha$, $\beta$ and $\gamma$ sheets, which are mainly derived from the Ru 4$d$-$t_{2g}$ orbitals. The $d_{yz}$- and the $d_{zx}$- orbitals give rise to the nearly one-dimensional $\alpha$-$\beta$-bands and the $ d_{xy} $-orbital yields the genuinely two-dimensional $\gamma$-band. 
Several studies suggest that while the $\gamma$-band is responsible for the superconductivity, the $\alpha$-$\beta$-bands rather contribute the enhanced incommensurate magnetic correlation~\cite{mazi97,ng00}. Note that strong incommensurate magnetic correlations are well known from neutron scattering experiments consistent with the simple band structure results~\cite{sidis99,braden02,braden04} and are also manifest as incommensurate magnetic order for Ti-doped ruthenates Sr$_2$Ru$_{1-x}$Ti$_x$O$_4$ with the $z$-axis polarization as expected theoretically from spin-orbit coupling~\cite{ti-braden02,ng00}.

In our previous study we restricted ourselves to a two-band model with the $d_{zx}$- and $d_{yz}$-orbitals, the $\alpha$-$\beta$-bands~\cite{imai12}. In this case edge states appear as Andreev bound states. As the topological number is zero due to the opposite chirality in the electron- and hole-like band there are two bound states that cross zero with opposite chirality. These states give rise to a net supercurrent flowing at the surface. Additionally we observe a spin current parallel to the surface,  which flows even in the normal state due to the structure of orbital hybridization and spin-orbit coupling. The spin orientation of the spin current is along the $z$-axis. The combination of super- and spin current leads to a net spin polarization. The almost flat dispersion of the Andreev bound states yields a large density of states at zero-energy. With repulsive interaction this can also lead to a spontaneous spin polarization at the surface due to the Stoner-like mechanism. Via spin-orbit and Hund's rule coupling the orientation of the supercurrent and the spin polarization are correlated. 

In the present study, we extend this analysis to all three bands which leads to a topologically non-trivial situation. We consider the $ \gamma $-band to be dominant and to carry a finite topological number. Including interactions and spin-orbit coupling of the three bands we examine the interplay between the chiral edge state derived from the $ \gamma $-band and the magnetism at the surface.

\section{Model and Method}
\begin{figure}[t]
\includegraphics[width=80mm]{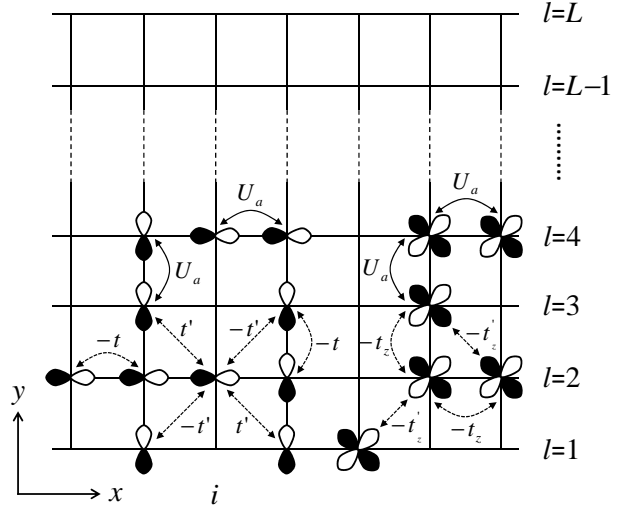}
\caption{Lattice structure of $L$-leg ribbon with three $d_{yx}$-, $d_{zx}$- and $d_{xy}$-orbitals. $t$ ($t'$) stands for the hopping integral between same (different) orbitals. $t_z$ ($t'_z$) stands for the hopping integral between $\gamma$-orbitals. $U_a$ represents the attractive interaction between the nearest neighbor sites. }
\label{lattice}
\end{figure}

First, we introduce the model Hamiltonian of the two-dimensional three-band system with ribbon shape. This geometry allows us to access the properties near the edges easier. This extended ladder-type system with the number of legs $L$ shown in Fig. \ref{lattice} has then two edges. 
We consider open boundary conditions in the $y$-direction leading to edge states. It is technically easy to turn to a bulk system by changing to periodic boundary conditions in this direction and choosing $L$ sufficiently large. We always assume translational invariance along the $x$-direction. The orbital structure of the 4$d$-$t_{2g}$ orbitals ($d_{zx}$, $d_{yz}$ and $d_{xy}$) suggests a specific pattern of nearest- and next-nearest neighbor hopping supplemented by onsite spin-orbit coupling, on the single-particle level. Moreover, we add onsite repulsive interaction, inter- and intra-orbital, including Hund's rule coupling. Nearest-neighbor interaction will be eventually used to introduce unconventional superconductivity in the spin-triplet channel. For a simple notation we label the orbitals $d_{zx}$, $d_{yz}$ and $d_{xy}$ hereafter by $x$, $y$ and $z$, respectively. 

Covering all these parts the Hamiltonian for the ribbon model can be written as
\begin{eqnarray}
H=H_{\alpha\beta}+H_{\gamma}+H_{\mu}+H_{\rm SO}+H_{\rm a}+H_{\rm r},
\label{eqn:ham}
\end{eqnarray}
with
\begin{eqnarray}
H_{\alpha\beta}&=&-t\sum_{i,\sigma}
\left(
 \sum_{l=1}^{L}c^{\dag}_{ilx\sigma}c_{i+1lx\sigma}+\sum_{l=1}^{L-1}c^{\dag}_{ily\sigma}c_{il+1y\sigma}
+ h.c.
\right)\nonumber\\
&&\hspace{-12mm}-t'\sum_{i,\sigma, m,m'\ne z}
 \sum_{l=1}^{L-1}\left(c^{\dag}_{ilm\sigma}c_{i+1l+1\bar{m}\sigma}-c^{\dag}_{il+1m\sigma}c_{i+1l\bar{m}\sigma}+ h.c.
\right),\\
H_{\gamma}&=&-t_z\sum_{i,\sigma}
\left(
 \sum_{l=1}^{L}c^{\dag}_{ilz\sigma}c_{i+1lz\sigma}+\sum_{l=1}^{L-1}c^{\dag}_{ilz\sigma}c_{il+1z\sigma}
+ h.c.
\right) \nonumber\\
&&\hspace{-8mm}-t_z'\sum_{i,\sigma}
 \sum_{l=1}^{L-1}\left(c^{\dag}_{ilz\sigma}c_{i+1l+1z\sigma}+c^{\dag}_{il+1z\sigma}c_{i+1lz\sigma}+ h.c.
\right), \\
H_{\mu}&=&-\mu\sum_{ilm\sigma}n_{ilm\sigma}-\Delta \varepsilon\sum_{il\sigma}n_{ilz\sigma},\\
H_{\rm SO}&=& -\lambda\sum_{il}
\sum_{mm'm''}\epsilon_{mm'm''}\sum_{\sigma \sigma'}c^{\dag}_{ilm\sigma}{\bm \sigma}_{\sigma \sigma' }^{m''}c_{ilm'\sigma'},\\
H_{\rm a}&=&
U_a\sum_{il\sigma \sigma'}\left(\sum_{\stackrel{m=}{x,z}}n_{ilm\sigma}n_{i+1lm\sigma'}+\sum_{\stackrel{m=}{y,z}}n_{ilm\sigma}n_{il+1m\sigma'}\right),
\label{hamatt}
\\
H_{\rm r}&=& U_r\sum_{ilm}n_{ilm\uparrow}n_{ilm\downarrow}
+K_r\sum_{ilm\ne m'}n_{ilm\uparrow}n_{ilm'\downarrow}\nonumber \\
&&+(K_r-J_r)\sum_{ilm<m'\sigma}n_{ilm\sigma}n_{ilm'\sigma},
\label{hamiltonian}
\end{eqnarray}
where $c^{\dag}_{ilm\sigma}$ ($c_{ilm\sigma}$) is the creation (annihilation) operator for electrons on the site $i$ of leg $l$, in the orbital $m$(=$x$-, $y$- or $z$-orbital) and with spin $\sigma$($=\uparrow$ or $\downarrow$). Moreover, $n_{ilm\sigma} =c^{\dag}_{ilm\sigma}c_{ilm\sigma}$ is the corresponding number operator and
 $\mu$ and $\lambda$ stand for the chemical potential and the amplitude of the spin-orbit coupling, respectively. The energy difference between the ($x$-$y$)-orbitals and $z$-orbital is denoted as $ \Delta \varepsilon $.    
$\epsilon_{mm'm''}$ and $\bm \sigma$ in $H_{\rm SO}$ are the Levi-Civita symbol and the Pauli matrix, respectively. Superconductivity is introduced by $ H_{\rm a}$ with $U_a$ as the attractive nearest neighbor interactions. Note the anisotropic structure for the $x$- and $y$-orbitals which only interact in the $x$- and $y$-direction, respectively.  
 The repulsive onsite interaction among the electrons includes intra-orbital $U_r$, inter-orbital $K_r$, and Hund's rule coupling $J_r$, which are important for the occurrence of magnetism near the edges. For simplicity, we ignored the exchange and pair hopping terms without changing the qualitative outcome. Note that we assume the standard relation $U_r = K_r + 2J_r$.

\begin{figure}[tb]
\includegraphics[width=50mm]{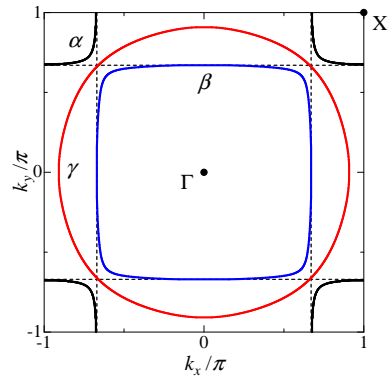}
\caption{(Color online) Fermi surfaces of two-dimensional bulk system for $t'=0.1t$, $t_z=0.7t$, $t'_z=0.3t$, $\Delta \varepsilon=0.065t$ and $\lambda=0.1t$ with $U_a=U_r=J_r=0$. 
The black, blue, and red lines stand for the $\alpha$-, $\beta$-, and $\gamma$-bands, respectively. The thin dashed lines stand for the Fermi surface of $\alpha$-$\beta$ bands without $t'$ and $\lambda$. }
\label{fs}
\end{figure}

To adjust the parameters, we examine the two-dimensional bulk Fermi surface for the non-interacting case ($U_a=U_r=J_r=0$) in Fig. \ref{fs} and compare them with the ones observed for Sr$_2$RuO$_4$. The two-dimensional bulk Hamiltonian can be written as
\begin{eqnarray}
H^{\rm 2D}&=&H_{\rm band}^{\rm 2D}+H_{\rm a}+H_{\rm r},
\label{eqn:2dham}
\\
H_{\rm band}^{\rm 2D}&=&\sum_{\bm k \sigma}\left\{\sum_{m}
 \varepsilon^m_{\bm k}c^{\dag}_{\bm k m \sigma}c_{\bm k m \sigma}
+\varepsilon^{xy}_{\bm k}\left(c^{\dag}_{\bm k x \sigma}c_{\bm k y \sigma}+h.c.\right)\right\}\nonumber\\
&+&H_{\rm SO}+H_{\mu},
\end{eqnarray}
where 
$\varepsilon^{x}_{\bm k}=-2t\cos k_{x}$, $\varepsilon^{y}_{\bm k}=-2t\cos k_{y}$, $\varepsilon^{xy}_{\bm k}=4t'\sin k_x \sin k_y$, and $\varepsilon^{z}_{\bm k}=-2t_z(\cos k_x+\cos k_y)-4t'_z\cos k_x \cos k_y$, and periodic boundary conditions are imposed in both directions. 
The $\alpha$- and $\beta$-bands, which mainly consist of the $x$- and $y$-orbitals have nearly one-dimensional characters with almost square-shaped Fermi surfaces.  If the orbital hybridization and spin-orbit interaction were absent, indeed only one-dimensional Fermi surfaces would appear (thin dashed lines in Fig. \ref{fs}). 
This leads to pronounced nesting features in the $\alpha$- and $\beta$-bands with the nesting vectors ${\bm Q}\sim (2\pi/3,0)$ or $(0,2\pi/3)$. 
The electron- (hole-) like Fermi surfaces are centered around the $\Gamma$ (X) point. 
The two-dimensional electron-like $\gamma$-band which consists of the $z$-orbital is connected with the $\alpha$-$\beta$-bands only through the spin-orbit coupling and the repulsive interaction. 

We introduce the BCS-type mean-field in the spin triplet channel to decouple the attractive interaction terms in $ H_{\rm a} $ in the usual way by the gap functions defined as 
\begin{eqnarray}
\Delta^{x}_{lm=x,z}&=&
\frac{1}{2}\left(\langle c_{i+1lm \uparrow}c_{ilm \downarrow}\rangle+\langle c_{i+1lm \downarrow}c_{ilm \uparrow}\rangle\right),\\
\Delta^{y}_{lm=y,z}&=&
\frac{1}{2}\left(\langle c_{il+1m \uparrow}c_{ilm \downarrow}\rangle+\langle c_{il+1m \downarrow}c_{ilm \uparrow}\rangle\right) , 
\end{eqnarray}
which corresponds to inplane equal-spin pairing. 
The components $\Delta^{x}_{ly}$ and $\Delta^{y}_{lx}$ vanish due to the anisotropic structure of the attractive interaction in Eq. (\ref{hamatt}). By its definition $\Delta^{y}_{lm}$ is not symmetric with respect to the center of the ribbon ($l=L/2$). However, the following relabeling yields a symmetric form, 
\begin{eqnarray}
\Delta^{y'}_{lm}&\equiv&
\left\{
\begin{array}{cc}
\Delta^{y}_{lm}/2&(l=1)\\
\Delta^{y}_{l-1m}/2&(l=L)\\
\left(\Delta^{y}_{l-1m}+\Delta^{y}_{lm}\right)/2&({\rm otherwise}).
\end{array}
\right.
\end{eqnarray}

The particle number and the spin polarization are defined as
\begin{eqnarray}
n_{lm}&=& n_{lm\uparrow}+ n_{lm\downarrow},\\
m_{lm}&=&n_{lm\uparrow}- n_{lm\downarrow},  
\end{eqnarray}
where $ n_{lm\sigma}\equiv \frac{1}{N}\sum_{i}\langle c^{\dag}_{ilm\sigma}c_{ilm\sigma} \rangle$. For the repulsive interaction terms we use the Hartree-Fock type mean-field decoupling leading to
\begin{eqnarray}
n_{ilm\sigma}n_{ilm'\sigma'}\rightarrow n_{lm\sigma} n_{ilm'\sigma'}+ n_{lm'\sigma'} n_{ilm\sigma}.
\end{eqnarray}

\section{Results}
\label{results}
We calculate the order parameters self-consistently with spatial resolution in the ribbon for $U_a =-1.5t$ at zero temperature. 
This leads to rather large gap functions such that the coherence length is short, only a few lattice constants. Hence, the number of legs $L=100$ is sufficient to ensure independent Andreev bound states at the two edges and the ribbon center displaying essentially bulk properties.  The other model parameters are chosen as follows: the particle number $n=4$, $t'=0.1t$, $t_z=0.7t$, $t'_z=0.3t$,  $\Delta \varepsilon=0.065t$ and $\lambda=0.1t$. 

\subsection{Superconducting property without repulsive interaction}
\begin{figure}[tb]
\includegraphics[width=50mm]{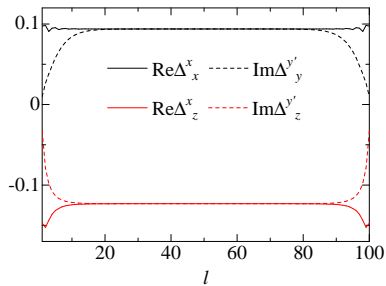}
\caption{(Color online) Gap functions as a function of leg index $l$ in $\bm d = \hat{z}(k_x+ik_y )$ superconducting state for $U_r=J_r=0$. 
}
\label{Delta1}
\end{figure}

We discuss first the superconducting order parameter in the absence of the repulsive interaction ($U_r=J_r=0$). We find that the most stable pairing state has the chiral $p$-wave form with $\bm d = \hat{z}(k_x+ik_y )$  avoiding nodes in the excitation gap, where the real and imaginary parts of the gap functions have relative phase difference ($\pi/2$) and same amplitudes (Re$\Delta^x$ = Im$\Delta^y$) in the bulk. Even in our ribbon model, this symmetry still remains at the center of the ribbon, as depicted in Fig. \ref{Delta1}. 

All gap functions, the one of the $\alpha$-$\beta$-bands ($\Delta^x_x$, $\Delta^y_y$) and the one of the $\gamma$-band ($\Delta^x_z$, $\Delta^y_z$), have same chirality, but opposite sign for the two band subsets.
Also in the ribbon model, the $\bm d = \hat{z}(k_x+ik_y)$ and $\bm d = \hat{z}(k_x-ik_y)$ state are degenerate and have the relation Re$\Delta^x=\pm $Im$\Delta^y$. Hereafter, we focus on $\bm d = \hat{z}(k_x+ik_y)$ state unless otherwise noted. 
 
\begin{figure}[!b]
\includegraphics[width=50mm]{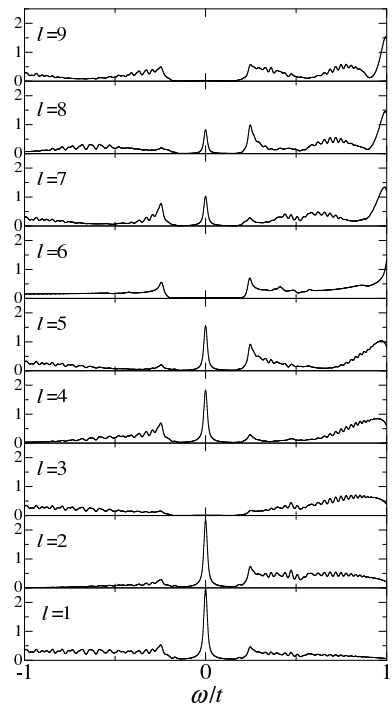}
\caption{
Spectral functions of the $y$-orbitals for various $l$ near an edge for $U_r=J_r=0$.}
\label{rho2}
\end{figure}

At the ribbon edges, $\Delta^y$  is suppressed and $\Delta^x$ is slightly enhanced~\cite{mats99}.  
Andreev bound states appearing at these edges give rise to the zero-energy peak in spectral function defined as
\begin{eqnarray}
\rho_{lm\sigma}(\omega)&=&\frac{1}{ N}\sum_{k,n}|u_{k}(lm\sigma,n)|^2\delta(\omega -E_{kn}),\\
\rho^{tot}_{l\sigma}(\omega)&=&\sum_{m}\rho_{lm\sigma}(\omega),
\end{eqnarray}
where $E_{k n}$ ($u_{k}(lm\sigma,n)$) is the energy eigenvalue (wave function eigenvectors) with the quantum number $n$ and the momentum $k$ of the mean-field Hamiltonian. 
\begin{figure}[tb]
\includegraphics[width=45mm]{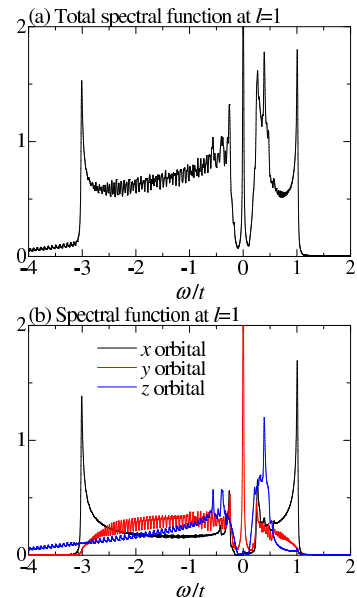}
\caption{(Color online) Spectral functions  at $l=1$ for $U_r=J_r=0$. (a) Total spectral function; (b) Each component of spectral function. 
}
\label{rho1}
\end{figure}
Figure \ref{rho1} shows the spectral functions near an edge for $U_r=J_r=0$. Note that the spectral function $\rho_{lm\sigma}(\omega)$ depends only very weakly on the spin index in the absence of the repulsive interaction. 
In the low-energy region, a rather sharp peak structure appears within the fully-opened superconducting gap of the chiral $p$-wave phase, as shown in Fig. \ref{rho1} (a).  Figure \ref{rho1} (b) indicates that the peak structure originates from the $y$-orbital, i.e. it is due to the Andreev bound state of the $p_y$-component of the pair wave function yielding the $\pi$-phase shift for quasiparticles bouncing off  the surface (see Fig. \ref{rho2} and discussion below). The contributions of the $x$- and $z$-orbitals are considerably smaller. Note that the one-dimensional character in the band related to the $x$-component still remains strongly visible through the singularities in the density of states at the band bottom and top. 

We now follow the zero-energy peak of the $y$-orbital as a function of the distance from the edge (see Fig. \ref{rho2}). With increasing distance $l$, the height of the peak is reduced gradually indicating that this peak represents a surface bound state.  
Note that the peak structure vanishes at $l=3\times$integer which indicates a spatial oscillation with the nesting wave vector $ Q \sim 2 \pi / 3 $, which corresponds to a Friedel-type of oscillation of the bound state wave function~\cite{tanu98,braun05}. 

In order to analyze the low-energy edge states, we compare the energy dispersions for the bulk and the ribbon model.  
\begin{figure}[!bt]
\includegraphics[width=50mm]{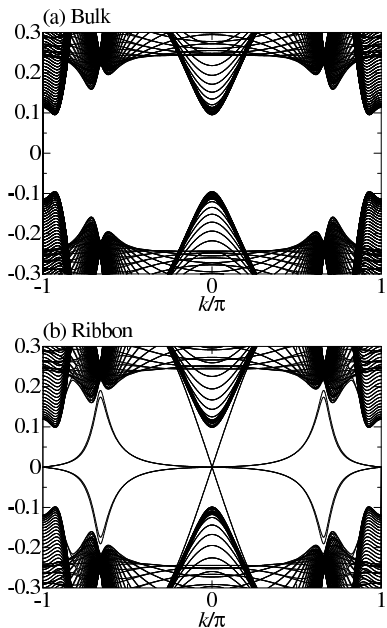}
\caption{Energy dispersion in the low-energy sector of the superconducting state; (a) bulk and (b) ribbon for $U_r=J_r=0$. 
}
\label{ek1}
\end{figure}
The chiral $p$-wave superconducting state  has by symmetry a nodeless quasiparticle gap in the bulk system, as shown in Fig. \ref{ek1} (a). 
Note that the Fourier transformation for the bulk and ribbon systems is introduced only in the $x$-direction as 
\begin{eqnarray}
c_{ilm\sigma}=\frac{1}{\sqrt{L_x}}\sum_{k}c_{klm\sigma}e^{-ikx_i}, 
\end{eqnarray}
where $k$ stands for the momentum along the $ x $-direction and $x_i$ is the $x$-coordinate of site $(i, l)$. $L_x$ is the number of sites to the  $x$-direction. 
The results for the bulk systems are taken by imposing periodic boundary conditions in the $y$-direction, such that for given $k$ as many energy levels
appear as in the ribbon case with open boundary conditions. 

In contrast to the bulk, the ribbon spectrum in Fig. \ref{ek1} (b) has subgap states which can be distinguished into two classes. There are those states forming almost flat bands with a strong dispersion only around $ k \sim 2 \pi / 3 $ and those states with a steep linear dispersion around $ k = 0 $. The former originate from the $\alpha$-$\beta$-bands. These flat bands are responsible for the larger peak around zero energy seen in Figs. \ref{rho1} and \ref{rho2}.
The latter are the chiral edge states of the $ \gamma $-band whereby each edge contributes one chiral branch which is topologically protected.

\subsection{Magnetic instability}

\begin{figure}[!tb]
\includegraphics[width=50mm]{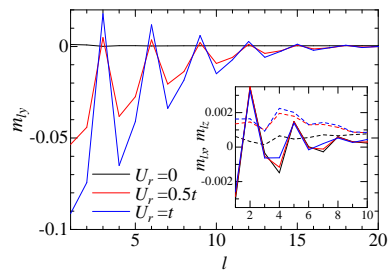}
\caption{(Color online) Spin polarization of the $y$-orbitals for several choices of $U_r$ near an edge. The inset shows spin polarizations of the $x$-orbitals (solid line) and $z$-orbitals (dashed line). }
\label{magur}
\end{figure}

We now turn on the repulsive onsite interactions which we keep small enough so as not to trigger an instability in the bulk system. Nevertheless, the spectral redistribution at the edges through the subgap states provides an environment where easily spin magnetism can emerge~\cite{imai12}. 
This becomes immediately obvious, if we look at the results for spontaneous spin polarization depicted in Fig. \ref{magur}, which is localized at the edges.  With increasing $U_r$, the amplitude of the spin polarization increases, but the penetration towards the bulk remains essentially unchanged. 
We notice that the spin polarizations of the $x$- and $z$-orbitals are almost independent of the repulsive interaction, while the $y$-orbital displays magnetism strongly depending on the repulsive interaction and, thus, dominating magnetism at the edges. 

\begin{figure}[!t]
\includegraphics[width=50mm]{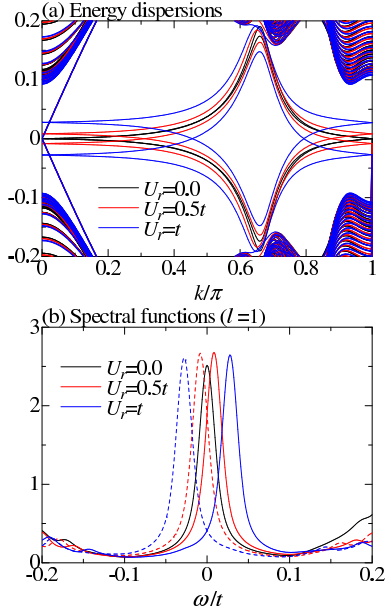}
\caption{(Color online) (a) Energy dispersions and (b) spectral functions at $l=1$ in the low-energy region for several choices of $U_r$ for $J_r=0$. The solid (dashed) line stands for the spectrum of the electron up (down) spin in lower panel. }
\label{ekrho}
\end{figure}

The magnetism which is present dominantly in the $x$-orbital due to the superposition of super- and spin current even for $U_r=0$ is much weaker and has only a minor contribution to the magnetism
despite transfer through next-nearest neighbor hopping and the spin-orbit coupling, as seen in the inset of Fig. \ref{magur}. This small spin polarization has, however, the role that it yields a bias for the spontaneous spin polarization of the $ y $-orbital. Thus, the magnetism is correlated with the orientation of the supercurrent. 
The oscillation of the spin polarization as seen in Fig. \ref{magur} is again of the type of a Friedel oscillation with the nesting wave vector $Q\sim2\pi/3$
and also indicates that the instability is basically triggered by the band structure nesting feature of the $\alpha$-$\beta$-bands. Figure \ref{ekrho} supports this 
observation as we can see that the spin polarization is due to the spin splitting of the essentially flat subgap states originating from the $\alpha$-$\beta$-bands
and the $ \gamma $-band subgap states are essentially unaffected. Thus, the Stoner-like spin instability is facilitated by the large density of states at zero energy provided by the near flat bands. 

\begin{figure}[b]
\includegraphics[width=50mm]{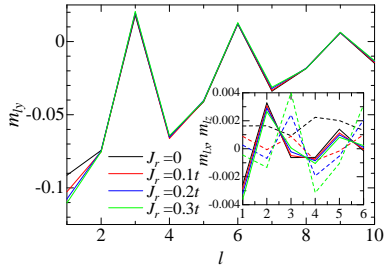}
\caption{(Color online) Spin polarization of the $y$-orbitals for several choices of $J_r$ near an edge for $U_r=t$. The inset shows spin polarizations of the $x$-orbital (solid line) and $z$-orbitals (dashed line). }
\label{mag_Jr}
\end{figure}

The effect of Hund's rule coupling $J_r$ is weak for the $ y $-orbital as well as the $ x $-orbital, but is rather influential for the $ z$-orbital, as can be seen in Fig. \ref{mag_Jr}. 

\begin{figure}[tb]
\includegraphics[width=50mm]{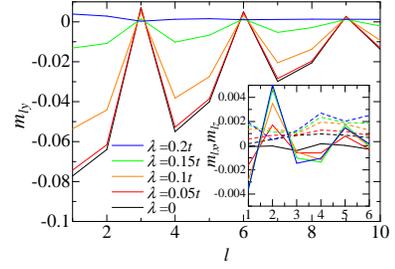}
\caption{(Color online) Spin polarization of the $y$-orbital for several choices of $\lambda$ near an edge for $U_r=0.5t$ and $J_r=0$. The inset shows spin polarizations of the $x$-orbital (solid line) and $z$-orbital (dashed line). }
\label{mag_lamba}
\end{figure}

The role of spin-orbit coupling is complex in view of the spontaneous magnetism of the $y$-orbital. On the one hand, it transfers the spin polarization between the different orbitals, but is also responsible for the spin currents at the surface, as mentioned earlier. Therefore, there is considerable effect due to spin-orbit coupling in the buildup of magnetism in both the $x$- and $z$-orbital as seen in the inset of Fig. \ref{mag_lamba}. On the other hand, spin-orbit coupling enhances the dispersion in the $ y $-orbitals such that its density of states at zero-energy is diminished leading to a reduction of the spin polarization, as can be observed in Fig. \ref{mag_lamba}.  

\subsection{Properties of charge and spin currents}
We now turn to the current densities, defining the spin-dependent current operators,
\begin{eqnarray}
j_{l\sigma}&\equiv& j^{\alpha \beta(1)}_{l\sigma}+j^{\alpha \beta(2)}_{l\sigma}+j^{\gamma(1)}_{l\sigma}+j^{\gamma(2)}_{l\sigma}, \nonumber \\
j^{\alpha \beta(1)}_{l\sigma}&=&\frac{1}{N}\sum_{k}(2t\sin k)c^{\dag}_{klx\sigma}c_{klx\sigma}, \nonumber \\
j^{\alpha \beta(2)}_{l\sigma}&=&\frac{1}{N}\sum_{m(=x,y)}
\left\{
 (-2{\rm i}t'\cos k)c^{\dag}_{klm\sigma}c_{kl+1\bar{m}\sigma}\right.\nonumber \\
&&\hspace{15mm}\left.+(2{\rm i}t'\cos k)c^{\dag}_{kl+1m\sigma}c_{kl\bar{m}\sigma}\right\},\\
j^{\gamma(1)}_{l\sigma}&=&\frac{1}{N}\sum_{k}(2t_z\sin k)c^{\dag}_{klz\sigma}c_{klz\sigma}, \nonumber \\
j^{\gamma(2)}_{l\sigma}&=&\frac{1}{N}\sum_{k}(2t'_z\sin k)(c^{\dag}_{klz\sigma}c_{kl+1z\sigma}+c^{\dag}_{kl+1z\sigma}c_{klz\sigma}).
\end{eqnarray}
Here we used the notation that $ \bar{m} $ means $ \bar{x} = y $ and $ \bar{y} = x $. 
Since $j^{n(2)}$ ($n=\alpha\beta$ or $\gamma$) in this form is not symmetric with respect to the center of the ribbon ($l=L/2$) we introduce the following redefinition for the purpose of display, 
\begin{eqnarray}
j^{n(2)'}_{l\sigma}&\equiv&
\left\{
\begin{array}{cc}
j^{n(2)}_{l\sigma}/2&(l=1)\\
j^{n(2)}_{l-1\sigma}/2&(l=L)\\
\left(j^{n(2)}_{l-1\sigma}+j^{n(2)}_{l\sigma}\right)/2&({\rm otherwise}).
\end{array}
\right.
\end{eqnarray}
Therefore, for the spin-dependent current density we use now, 
\begin{eqnarray}
j^{n}_{l\sigma}&\equiv& j^{n(1)}_{l\sigma}+j^{n(2)'}_{l\sigma}. 
\end{eqnarray}

\begin{figure}[!tb]
\includegraphics[width=50mm]{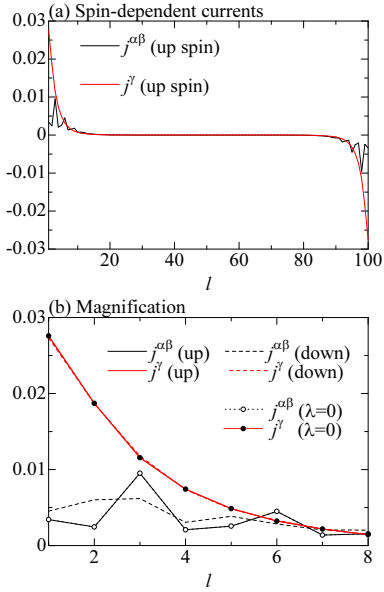}
\caption{(Color online) (a) Spin $\uparrow$ currents for $\bm d=\hat{z}(k_x+ik_y)$ state as a function of leg index $l$ for $\lambda=0.1t$ and $U_r=J_r=0$. (b) Magnification of (a); Solid lines stand for spin $\uparrow$ and $\downarrow$ electron for $\lambda=0.1t$, and open (closed) circle represents the current for $\lambda=0$. }
\label{current}
\end{figure}
In the absence of the repulsive interactions, the spin-dependent currents from the $\alpha$-$\beta$- and the $\gamma$-bands along the $x$-direction are depicted in Fig. \ref{current}.  
The spin-dependent currents appear near the edges, and the sum of each current $j_{l\sigma}$ becomes the net charge current flowing along the edges. Because of time-reversal symmetry breaking in the superconducting state, both the spin $\uparrow$ and $\downarrow$ currents from all $\alpha$-$\beta$- and $\gamma$-bands flow in the same direction. The flow direction and the amplitudes are associated with the sign of the chirality or of the spin-orbit interaction, in which the following relations have to be satisfied,
\begin{eqnarray}
j^{n}_{l\sigma} \,(\bm d=\hat{z}(k_x+ik_y))&=&-j^{n}_{l\bar{\sigma}} \,(\bm d=\hat{z}(k_x-ik_y)), \\
j^{n}_{l\sigma} \,({\rm sign}\lambda=+1)&=&j^{n}_{l\bar{\sigma}} \,({\rm sign}\lambda=-1). 
\end{eqnarray}
With the gap functions of both the $\alpha$-$\beta$- and $\gamma$-bands having same chirality, the flow directions of the currents are the same. $ j^{n(1)}_{l\sigma}$ component is larger than that of $ j^{n(2)}_{l\sigma}$, and is dominant in the current, which indicates that in contrast to the magnetic property, the currents  are dominated by the $x$- and $z$-orbitals. 

The spin-orbit interaction yields  a difference between the spin $\uparrow$ and $\downarrow$ currents, as has been discussed in Ref.~\cite{imai12} for the
$\alpha$-$\beta$-bands. There the combination of hopping and spin-orbit coupling yields circular spin current patterns which cancel in the bulk but yield a net spin currents at the surface due to the lack of cancellation. Spin-orbit coupling together with the chiral supercurrents leads also to a weak spin current in the $ \gamma $-band.  

\begin{figure}[!tb]
\includegraphics[width=50mm]{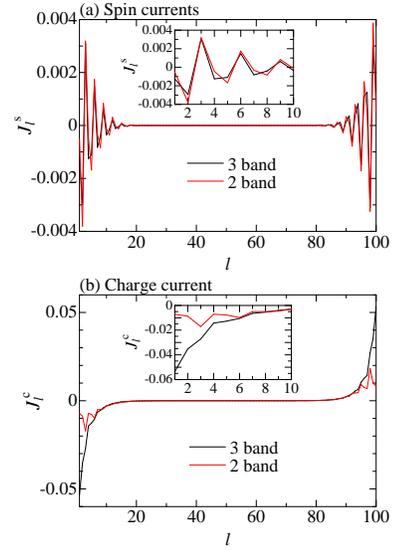}
\caption{(Color online) Charge and spin currents as a function of leg index $l$ for $U_r=J_r=0$.  The insets show the magnifications near an edge. }
\label{currentsc}
\end{figure}

\begin{figure}[b]
\includegraphics[width=50mm]{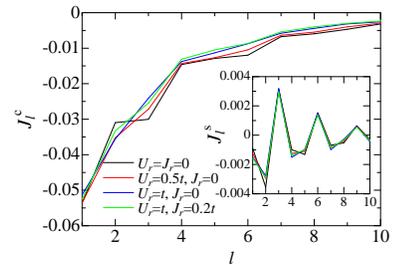}
\caption{(Color online) Charge current as a function of leg index $l$ near an edge for several choices of $U_r$ and $J_r$. The inset shows spin current. }
\label{currentsc_Ur}
\end{figure}

Now let us compare the charge and spin currents defined as
\begin{eqnarray}
\hat{J}^c_l&=&-\frac{e}{\hbar}\sum_{\sigma}(j^{\alpha \beta}_{l\sigma}+j^{\gamma}_{l\sigma}),\\
\hat{J}^s_l&=& \sum_{\sigma}\sigma(j^{\alpha \beta}_{l\sigma}+j^{\gamma}_{l\sigma}),
\end{eqnarray}
for the two-band model~\cite{imai12} and the three-band model. Figure \ref{currentsc} shows that essential differences can be found largely in the charge but much less in the spin current distribution, in the absence of repulsive interactions. 
Note that Fig. \ref{current} (b) shows that the $\gamma$-band does not carry spin currents in this case. 
The orientation of the charge current remains the same, but the magnitude is considerably larger for the three band model due to the contributions of the $\gamma $-band. The lack of change for the spin current is not surprising in view of the mechanism driving the spin currents in the $\alpha$-$\beta$-bands. Interaction effects are not strong on both current densities, as Fig. \ref{currentsc_Ur} shows. 

\subsection{Induced magnetic fields}
\begin{figure}[!tb]
\includegraphics[width=50mm]{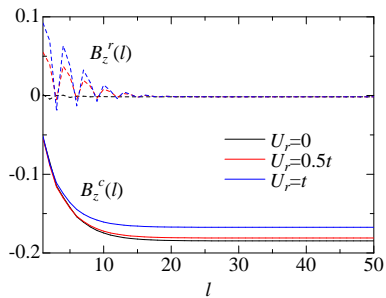}
\caption{(Color online) Spontaneous magnetic field for several choices of $U_r$ for $\lambda=0.1$ and $J_r=0$. The solid (dashed) line stands for  $B_z^r(l)$ ($B_z^c(l)$).   }
\label{smf}
\end{figure}

Both the charge current and the spin polarization yield the net magnetic field. By means of the Maxwell's equation $\nabla \times \bm B =\mu_0 \bm j$, we obtain the magnetic field from the charge current, which is given by
\begin{eqnarray}
B_z^c(l)=\mu_0 \sum_{l'}^{l}\langle J^c_{l'}\rangle=-\frac{et}{\hbar a} \mu_0\sum_{l'}^{l}\langle \tilde{J}^c_{l'}\rangle, 
\end{eqnarray}
where $\langle \tilde{J}^c_{l'}\rangle$ represents the dimensionless current density. 
On the other hand, the spin polarization generates the following magnetic field, 
\begin{eqnarray}
B_z^r(l)=-\frac{\mu_0}{a^2c}\mu_{\rm B}(n_{l \uparrow}-n_{l \downarrow}), 
\end{eqnarray}
where $\mu_{\rm B}$ is the Bohr magneton and  $a$ ($c$) is the lattice constant in the $x$- and $y$- ($z$-) direction. We stress that both prefactors in $B_c^r(l)$ and $B_z^r(l)$ have the same order of magnitude. 
Figure \ref{smf} shows  these two fields without the prefactors, respectively. 
Here Meissner-Ochsenfeld screening effects are not taken into account, which would limit the current-induced field to the surface region. 

Although the presence of the $\gamma$-band increases the magnetic field due to the charge current in comparison with the case in the two-band model, both magnetic fields are still of comparable magnitude and opposite sign. Therefore the net magnetic field can be reduced due to the compensation, which leads to a reduction of the overall field that could be measured at the edges, as already discussed in Ref.~\cite{imai12}.

\subsection{Topological property}

Finally, we address the topological properties of the superconducting phase in the three-band model. For the materials with the fully-opened insulating or superconducting gap, the edge states are closely related with topological properties of the bulk state due to the bulk-edge correspondence~\cite{wen92,hats93}.
In order to study the topological properties, we consider the two-dimensional bulk Hamiltonian given by  Eq. (\ref{eqn:2dham}), including the BCS-like decoupling of the pairing interaction like in the ribbon model. 

We define the topological invariant in the superconducting phase, following Ref. ~\cite{volo89}, as 
\begin{eqnarray}
N_T&=&-\frac{2\pi i}{N}\sum_{\bm k}\sum_{ijkl}\sum_{n,n'}
(J_{\bm k}^{x})_{ij}(J_{\bm k}^{y})_{kl}\nonumber \\
&&\hspace{-15mm}\times u_{\bm k}^*(i,n)u_{\bm k}(j,n')
u_{\bm k}^*(k,n')u_{\bm k}(l,n)
\frac{f(E_{\bm kn})-f(E_{\bm kn'})}{(E_{\bm kn}-E_{\bm kn'})^2},
\label{eq:Hall}
\end{eqnarray}
where indices $i, j, k, l$ include the site, orbital and spin. 
The matrix element $J_{\bm k}^{\mu}$ is formally defined like a current, 
\begin{eqnarray}
(J_{\bm k}^{\mu})_{ij}=\langle i| \frac{\partial H_{\bm k}}{ \partial k_\mu}| j\rangle, 
\label{eq:jk}
\end{eqnarray}
with the mean-field Hamiltonian $H^{2D}_{MF}=\sum_{\bm k} H_{\bm k}$, 
 where $E_n$ stands for the energy eigenvalue of the quasiparticle state with index $n$ in the two-dimensional bulk system and $f(E_n)$ is the Fermi distribution function. 
Note that in the superconducting phase, Eq. (\ref{eq:jk}) does not have the meaning of physical charge current~\cite{lutc09}. 
$N_T$ becomes an integer and corresponds to so-called Chern number in case of a topologically non-trivial state. 

\begin{figure}[b]
\includegraphics[width=50mm]{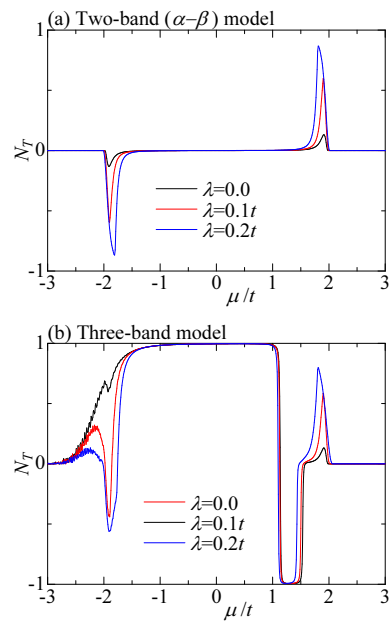}
\caption{(Color online) Topological number as a function of chemical potential $\mu$ for several choices of $\lambda$ with $\bm d=\hat{z}(k_x+ik_y)$: (a) two- and (b) three-band models. }
\label{Chern}
\end{figure}
Figure~\ref{Chern} shows the topological number $ N_T$  as a function of the chemical potential, where clear difference between the two-band and three-band model can be observed. 
Note that the electron number and the Fermi surface structure of Sr$_2$RuO$_4$ are reproduced at $\mu \approx t$. 
In the two-band model, there are two Fermi surfaces, i.e. an electron- and a hole-like ones, whose Chern numbers are opposite and cancel perfectly. 
Thus, $N_T$ vanishes in the wide range of $\mu$ shown in Fig.~\ref{Chern} (a). 
The finite $N_T$, however, appears only near the bottom or top of the two bands, where either the electron or the hole Fermi surface is completely depleted.
The range of non-zero $N_T$ can be enlarged by increasing spin-orbit coupling which affects the band structure. 
The reason why $ N_T $ does not reach an integer value can be attributed to the fact that the Fermi surface is very close to a symmetry point in the Brillouin zone: the $ \Gamma $-point for $ \mu \approx -2t $ and the X-point for $ \mu \approx +2t $, where the gap function turns to be zero.  
Thus, the condition for topological protection is not given in the two-band model because the Fermi surface in the superconducting phase is blurred.

\begin{figure}[tb]
\includegraphics[width=50mm]{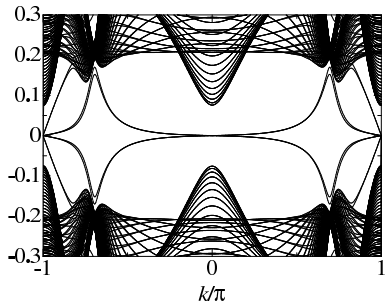}
\caption{Energy dispersion in the low-energy sector of the superconducting state at $\mu=1.2t$.}
\label{mu120}
\end{figure}
Since the two-band system has vanishing Chern number in the wide range of $ \mu$, the observed finite $N_T$ in the three-band model originates from the $ \gamma $-band, which 
provides an integer value $ N_T = +1 $ or $-1$, as shown in Fig.~\ref{Chern} (b).
In addition, an interesting aspect arises, i.e. the sign change of $ N_T $ at $\mu=\mu_c \approx 1.15t$.  
This indicates that the $ \gamma $-band shows a Lifshitz transition from electron- ($\mu < \mu_c $) to hole-like ($\mu > \mu_c $). 
Here the topological sector is switched and the chiral edge states shift from crossing zero at $ k_x = 0 $ to $ k=\pm \pi $.
Figure~\ref{mu120} shows the energy dispersion at $\mu=+1.2t$, where the Fermi surface touches the X-point resulting in vanishing of the gap function due to the symmetry. 
Since the chemical potential of Sr$_2$RuO$_4$ is rather close to $ \mu_c $, the chiral edge state might be fragile against disorder~\cite{FRG13}.

We would like to comment on the initial discrepancy in the topology of the $ \gamma $-Fermi surface in ARPES measurements. 
Early ARPES results indicated a hole-like $ \gamma $-sheet~\cite{yokoya96}, while later experiments confirmed an electron-like Fermi surface consistent with de Haas-van Alphen measurements~\cite{dama00}. 
The position of the Fermi level for the $ \gamma $-band near the $z$-axis oriented surface may be subtle and depend on surface state properties. 
Moreover surface reconstruction doubling the unit cell, affecting particularly the $ z$-orbital, has been reported which complicates the topology of the superconducting phase on the $\gamma$-band~\cite{matz00}. Thus, it would not be surprising that near the $z$-oriented surfaces the chiral edge states may become rather fragile to disorder. This aspect of the $\gamma$-band will be discussed elsewhere in more detail.

\section{Summary and Discussions}
\label{summary}

We have examined the complex interplay between the topological aspects of the superconducting phase, band structure driven surface spin currents and correlation-driven magnetism for a three-band model corresponding to Sr$_2$RuO$_4$. In a model favoring spin-triplet pairing a chiral $p$-wave superconducting phase is realized which appears with the same chirality coupled in all three bands ($\alpha$-, $ \beta$- and $ \gamma $-band) connected through spin-orbit coupling and inter-band hybridization. The topological nature of the chiral $p$-wave state is visible in the $ \gamma $-band which gives rise to a chiral edge mode following the expectations. The $ \alpha $-$\beta$-bands together give also rise to  Andreev bound states at the surface, which are not topologically protected. However, due to the nearly flat dispersion these states generate a large zero-energy density of states, which make them  very susceptible to a Stoner-like magnetic instability. Note that this is compatible with magnetic correlations due to the nesting feature of the $ \alpha$-$\beta$-bands which leads to strong incommensurate magnetic correlations as observed by neutron scattering~\cite{sidis99}. 

We would like to emphasize here the connection between the correlation-driven magnetism and the chiral superconducting phase. This is caused by spin polarization induced by the combination of normal state spin currents at the surface carried by the $ \alpha$-$\beta$-band electrons and the chiral edge states. This correlation of chiral and spin edge magnetism may give rise to a cancellation of the magnetic signal at the surface, which could be an explanation for the negative result in the search for chiral magnetism in Sr$_2$RuO$_4$. 

Note that the spin-orbit coupling between $ \alpha $- and $ \beta $-bands acquires Rashba-type features near the surface, since the hybridization between different orbitals lacks inversion symmetry. The $ \gamma $-band does not have this feature. Naturally there is a slight modification of the orbital structures at the surface such that additional Rashba-like spin-orbit coupling appears. This is, however, considerably weaker as it involves hybridizations which are completely suppressed in the bulk by symmetry. Modification due to these corrections are small, e.g. changing slightly the surface density of states at zero energy, which would affect the magnetic instability slightly. However, no qualitative change appears. 

Another aspect which is of interest in the context of edge state is the topological nature of the superconducting phase. By evaluating the Chern number we show that the $ \gamma $-band is intrinsically a topologically non-trivial phase, while the $ \alpha$-$\beta$-bands constitute together in the relevant 
band-filling range a topologically trivial subsystem. However, a closer look at the Chern number shows that the actual system is close to a Lifshitz transition which would yield a switch of the sign of the Chern number. While this does not affect the edge states essentially in a clean system, it potentially makes them fragile to disorder. Also this is a feature which can jeopardize the observation of chiral edge currents in the system. It also may provide opportunities for novel phenomena in a topological superconductor, if Sr$_2$RuO$_4$ turns out to be in this class.

\begin{acknowledgments}
We are grateful to A. Bouhon, T. Neupert, T.M. Rice, and T. Saso.
The work is partly supported by Swiss Nationalfonds and the NCCR MaNEP through the special project ``Topomatter''. K. W. acknowledges the financial support  of KAKENHI (No. 23310083). 
\end{acknowledgments}

\appendix

\bibliography{paper.bib}

\end{document}